\documentclass[aps,footinbib,twocolumn,showpacs,amsmath,amssymb,superscriptaddress,prb,groupedaddress]{revtex4}
\usepackage{graphicx,amsmath}
\usepackage{epstopdf}
\usepackage{amssymb}
\usepackage{grffile}
\usepackage[usenames]{color}
\usepackage{indentfirst}
\usepackage{float}
\usepackage{color}
\usepackage{mathrsfs}
\usepackage{dcolumn}
\usepackage{bm}
\usepackage[colorlinks=true,bookmarks=false,citecolor=blue,linkcolor=red,urlcolor=blue]{hyperref}
 
\def\red{\textcolor{red}}

\begin{document}
\title{Anyonic braiding via quench dynamics in fractional quantum Hall liquids}
\author{Jie Li$^{a}$}
\affiliation{Department of Physics and Chongqing Key Laboratory for Strongly Coupled Physics, Chongqing University, Chongqing 401331, People's Republic of China}
\author{Dan Ye$^{a}$}
\thanks{$^{a}$These authors have contributed equally to this work.}
\affiliation{Department of Physics and Chongqing Key Laboratory for Strongly Coupled Physics, Chongqing University, Chongqing 401331, People's Republic of China}
\author{Chen-Xin Jiang}
\affiliation{Department of Physics and Chongqing Key Laboratory for Strongly Coupled Physics, Chongqing University, Chongqing 401331, People's Republic of China}
\author{Na Jiang}
\affiliation{School of Materials Science and engineering, Chongqing Jiaotong University, Chongqing 400074, People's Republic of China}
\author{Xin Wan}
\affiliation{Zhejiang Institute of Modern Physics, Zhejiang University, Hangzhou 310027, People's Republic of China}
\affiliation{CAS Center for Excellence in Topological Quantum Computation, University of Chinese Academy of Sciences, Beijing 100190, People's Republic of China}
\author{Zi-Xiang Hu}
\email{zxhu@cqu.edu.cn}
\affiliation{Department of Physics and Chongqing Key Laboratory for Strongly Coupled Physics, Chongqing University, Chongqing 401331, People's Republic of China}

\begin{abstract}
In a Laughlin fractional quantum Hall state, one- and two-quasihole states can be obtained by diagonalizing the many-body Hamiltonian with a trapping potential or, for larger systems, from the linear combination of the edge Jack polynomials. The quasihole states live entirely in the subspace of the lowest-energy branch in the energy spectrum with a fixed number of orbits, or a hard-wall confinement. The reduction in the Hilbert space dimension facilitates the study of time evolution of the quasihole states after, say, the removal of the trapping potential. We explore the quench dynamics under a harmonic external potential, which rotates the quasiholes in the droplet, and discuss the effect of long-range interaction and more realistic confinement. Accurate evaluation of the mutual statistics phase of anyons for a wide range of anyon separation can be achieved from the Berry-phase calculation.
\end{abstract}
\date{\today}
\maketitle

 \section{Introduction}
A remarkable feature of the fractional quantum Hall (FQH) effect~\cite{Tsui} is the emergence of
fractionally charged excitations with exotic statistics, dubbed anyons.~\cite{Laughlin83,Wilczek,Halperin,Arovas,WuYS}
The presence of these collective excitations is a manifestation of the underlying topological phases.~\cite{Wenbook}
Abelian anyons are so named because their interchanges lead to an Abelian phase factor $e^{i\theta}$, where $\theta$ can be neither an integral multiple of $2\pi$ for bosons nor an odd multiple of $\pi$ for fermions.
Non-Abelian anyons, on the other hand, live in a degenerate ground-state manifold of topological origin.
When they braid around each other, their wave function undergoes a unitary transformation in the degenerate space.~\cite{MR91}
Their potential application as resources in topological quantum computation has aroused great interest in recent years.~\cite{Kitaev, NayakRMP}

In theory, the property of fractional charge manifests in the Aharonov-Bohm phase
acquired by a single anyon along a closed path that enclosed no anyons.
The fractional exchange properties, however, arises in the mutual statistical part
of the Berry phase, when additional anyons are present in the loop.~\cite{Arovas}
So far, experimental evidences mounts for fractional charge of both Abelian (i.e., $e/3$ in a $\nu=1/3$ Laughlin-like state) and, presumably, non-Abelian anyons (i.e., $e/4$ in a $\nu=5/2$ Moore-Read--like state).~\cite{Goldman,Mahalu,Etienne,Yacoby, Mahalu08,Radu}
However, an unambiguous experimental demonstration of fractional statistics has been under intense debate.
Inconsistencies between the experimental results and theoretical predictions exist in Fabry-P\'erot interferometers~\cite{Mahalu10, West12, Umansky, Willett} or Mach-Zehnder interferometers.~\cite{Ji03, Mailly}
The obstacles in observing the fractional statistics include the Coulomb blockade effect in a small interferometer~\cite{Halperin11} and the decoherence due to neutral modes in a large interferometer.~\cite{Umansky19}

In novel devices with screening layers conductance oscillations have been observed
with abrupt jumps in the interference pattens consistent with the theoretically expected anyonic statistics.~\cite{Sahasrabudhe,NakamuraNP19,NakamuraNP20,WhitNC}
Moreover, the fractional statistics of FQH quasiholes has also been observed
in the current correlation resulting from the collisions among anyons at a beam splitter.~\cite{BartolomeiScience}
The achievements heighten the need to study controllable anyon braiding toward the implementation of topological quantum computation.
In general, anyons can be created and trapped in the bulk, but dragging them with scanning tunneling microscope (STM) tips is far from reality.
On the other hand, in a Fabry-P\'erot or Mach-Zehnder interferometer, anyons propagate along sample edges with a finite velocity given by the dispersion of gapless edge
modes.~\cite{Wen1,Wen2, Wan03, Hu09}
However, one cannot ignore the details of the edge potential profile in time evolution.

In this work, we construct one- and two-quasihole wave functions in a Laughlin FQH system by introducing an appropriate trapping potential in the interacting Hamiltonian or through linear combinations of Jack polynomials for edge modes.
We realize anyonic braiding in the bulk in the quench dynamics after we suddenly remove the tip potential. The projection of the quasihole states to edge wave functions helps us reduce the Hilbert space tremendously in the time evolution.
We can introduce an external harmonic potential to confine the anyon excitations and to tune the braiding period, or explore the smear-out of the anyons and their revival in the presence of long-range interaction.  
We also verify the fractional statistics by calculating the Berry phase in a period, which can be related to 
the mean angular momentum of the states with various number of quasiholes. 

The rest of the paper is organized as follows.
In Sec.~\ref{sec:qh} we describe how to construct quasihole states with model Hamiltonian or with edge Jack polynomials.
We demonstrate in Sec.~\ref{sec:fs} the fractional statistics by calculating the mean angular momentum, or equivalently, the Berry phase in the wave-function evolution.
In Sec.~\ref{sec:quench} we present a quench protocol to realize the quasihole braiding and discuss the effect of long-range interaction and edge confinement. We conclude in Sec.~\ref{sec:dis} and present additional details on constructing quasihole states with Jack polynomials in Appendix ~\ref{sec:app_A}.

\section{Preparing Quasihole states}
\label{sec:qh}

Consider a $\nu = 1/3$ Laughlin FQH droplet with a quasihole state at $w$ in a complex plane
with model wave function~\cite{Laughlin83}
 \begin{eqnarray}
  \Psi_{\text{L}}^{\text{1QH}} = \prod_i (z_i - w)\Psi_{\text{L}}
 \end{eqnarray}
where
\begin{equation}
\Psi_{\text{L}} = \prod_{i<j}(z_i - z_j)^3 e^{-\sum_i |z_i|^2/4l_B^2}
\end{equation}
is the Laughlin wave function with a total angular momentum $M_0 = 3N_e(N_e-1)/2$
and $z_k = x_k + i y_k$ is the complex coordinate for the $k$th electron.
And $l_B$ is the magnetic length.
A two-quasihole wave function can be written as
\begin{eqnarray}
 \Psi_{\text{L}}^{\text{2QH}} = \prod_i(z_i - w_1)\prod_i(z_i - w_2)\Psi_{\text{L}}
\end{eqnarray}
with $w_1$ and $w_2$ being the positions of the quasiholes.
In the pseudopotential formalism,~\cite{HaldanePP} $\Psi_{\text{L}}$ is the densest zero energy eigenstate for the $V_1$-only Hamiltonian.
In the interacting Hamiltonian, a trapping potential, which models the effects of an STM tip or a localized impurity, can be introduced to confine a quasihole in the bulk.
A previous study~\cite{Hu08} showed that a repulsive Gaussian potential
\begin{equation}
V(z) = \frac{U}{\sqrt{2 \pi} s} \exp[-\vert z - w \vert^2/(2s^2)]
\end{equation}
can be used to trap a quasihole at $w = w_r + i w_i$.
Here, $U$ and $s$ are the strength and width of the trapping potential.
In the lowest Landau level with the symmetric gauge, the rotationally invariant basis is
\begin{equation}
|m\rangle = \frac{1}{\sqrt{2\pi}l_B} \frac{1}{\sqrt{2^m m!}} z^m e^{-|z|^2/4l_B^2}.
\end{equation}
and the corresponding matrix elements are $V_{mn} = \langle m | V(z) | n \rangle$.
Non-diagonal terms appear when the rotational symmetry is broken by $w \neq 0$.
Without loss of generality, we assume $q \equiv n-m > 0$
and obtain the integral form of the potential matrix elements as
 \begin{eqnarray}
 V_{mn} &=& \frac{U s}{(2\pi)^{1/2}\sqrt{2^{q}m!}}\frac{\sqrt{n!}}{q!}\int_0^{\infty} dk e^{-\frac{k^2s^2}{2}} k^{q+1} \nonumber \\
  &&\times _1F_1 \left [n+1, q+1, -\frac{k^2}{2} \right ] J_{q} (k \vert w \vert) \left ( \frac{w}{\vert w \vert} \right )^{-q},
 \nonumber
 \end{eqnarray}
where $_1F_1$ and $J_{n-m}$ are the confluent hypergeometric function and Bessel function, respectively. In the limit of zero width $s \rightarrow 0$, the Gaussian can be replaced by a $\delta$-function $V(\vec{r}) = U_\delta \delta(z - w)$,
and the matrix elements are simplified to
  \begin{eqnarray}
   V_{mn} &=& \frac{U_\delta}{2^{q/2}} \sqrt{\frac{m!}{n!}} \int_0^{\infty} dk e^{-\frac{k^2}{2}} k^{q+1}
   L_{m}^{q} \left (\frac{k^2}{2} \right ) \nonumber \\
   &&\times J_{q}(k \vert w \vert) \left ( \frac{w}{\vert w \vert} \right )^{-q}.
  \end{eqnarray}
In practice, we use the Gaussian potential with width $s \simeq 2l_B$ for the full Coulomb interaction and the $\delta$-potential for the $V_1$-only interaction to create a quasihole.~\cite{Hu08}

To simulate the braiding of two quasiholes, we must have a sufficient large system
to avoid quasihole overlap.
However, large systems are difficult to solve by exact diagonalization
due to the absence of rotational symmetry.
We, therefore, introduce a truncated subspace for finite systems in disk geometry
with the help of Jack polynomials, such that we can calculate up
to $12$ electrons at $1/3$ filling.

\begin{figure}
	\includegraphics[width=9cm]{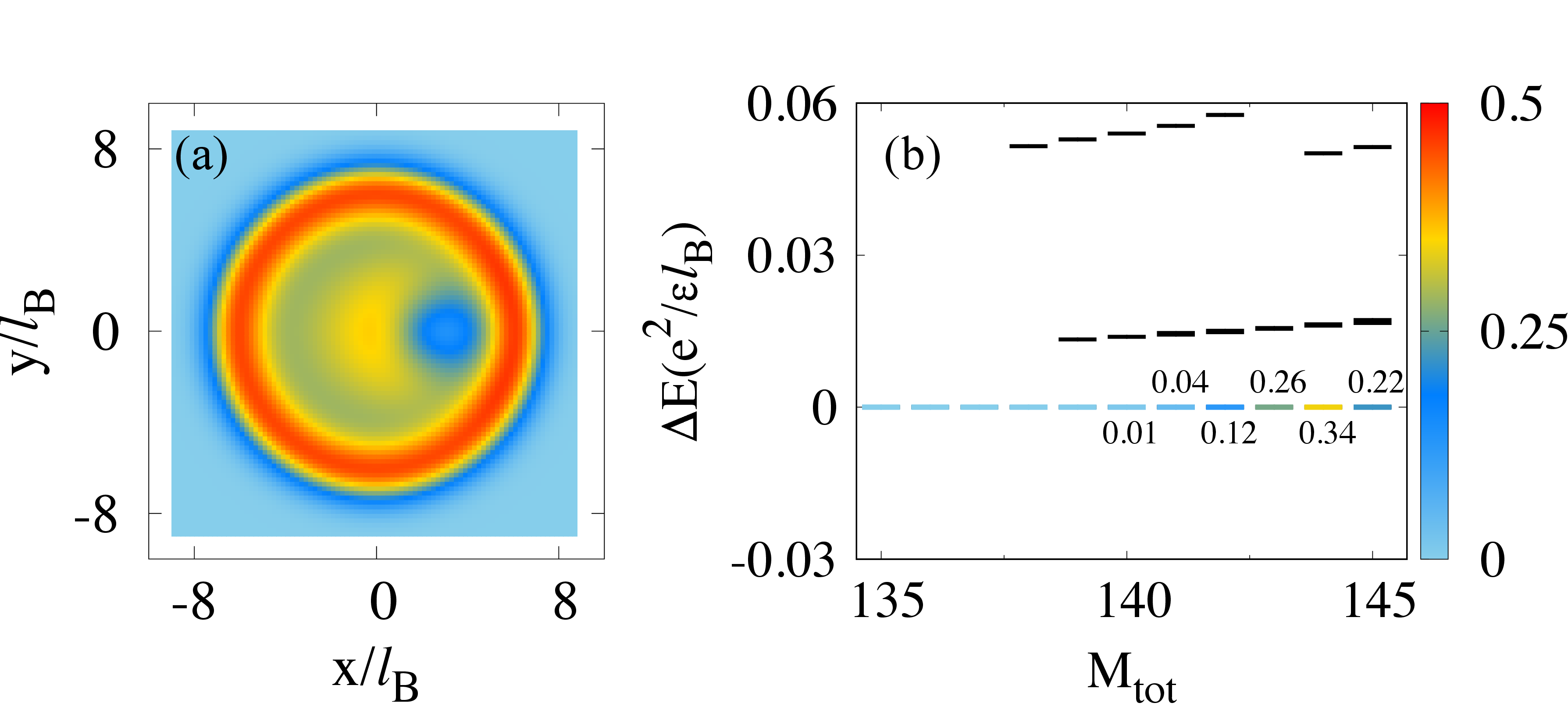}
	\caption{\label{oneqh}(a) The one-quasihole density profile for 10 electrons with a quasihole at $w = 3l_B$. (b) The low-lying energy spectrum for the 10 electrons in 29 orbitals. The color illustrates the overlap of these edge states with the one-quasihole model wave function.}
\end{figure}

Jack polynomials, or Jacks, are homogeneous symmetric polynomials specified by a rational parameter $\alpha$ and a root configuration $\mu$.
Up to the Gaussian factor, FQH model wave functions can often be expressed exactly as Jacks.~\cite{Bernevig1,Bernevig2}
We denote the Jack with the root configuration $\mu$ by $\vert \mu \rangle$ in the following,
while we have a fixed $\alpha = -2$ for the $\nu = 1/3$ Laughlin state
and its quasihole and edge excitations.
The extra polynomial corresponding to the quasihole excitation can be expanded to be
\begin{eqnarray}
\prod_i(z_i - w) &=& \prod_i z_i - w \sum_{j} \prod_{i\neq j} z_i + w^2 \sum_{j,k} \prod_{i \neq j, k} z_i \nonumber \\
&+& \cdots + (-1)^{N_e}w^{N_e},
\end{eqnarray}
where $N_e$ is the number of electrons.
The $n$th term on the right-hand side is a symmetric polynomial of $z_i$s of order $N_{e}-n + 1$.
In general, $\Psi_{\text{L}}$ multiplied by a symmetric polynomial can be interpreted
as an edge excitation.~\cite{LeeJack, Hu16}
For example, $\sum_{ij}z_i z_j$ and $\sum_i z_i^2$ span the edge space
with angular momentum $\Delta M_{tot} = M_0 + 2$.
In fact, each term in the above expansion is the unique (zero-energy) ground state of
the $V_1$ Hamiltonian in their own momentum subspace for $N_e$ electrons in $3N_e-1$ orbitals,
which corresponds to a single Jack.
So we can rewrite the one-quasihole state as
 \begin{eqnarray}\label{onejack}
 \Psi_{\text{L}}^{\text{1QH}} &=& |01001001\cdots 1001\rangle - w |10001001\cdots 1001\rangle \nonumber \\
 &+& w^2 |10010001\cdots 1001\rangle + \cdots \nonumber \\
 &+& (-1)^{N_e} w^{N_e}|1001\cdots 10010\rangle,
\end{eqnarray}
where the notation $|1001\cdots\rangle$ represents an unnormalized Jack polynomial and we neglect the Gaussian factor for convenience.
In Fig. ~\ref{oneqh}(a), we plot the density profile for a $N_e = 10$ state with one quasihole at $w = 3l_B$ on the real axis.
The wave functions obtained from diagonalizing the $V_1$-only Hamiltonian with the $\delta$-potential at $w$ and from the construction by Jacks are identical.
To illustrate how this quasihole state can be decomposed into a series of edge states,  
we plot the low-energy spectrum for the $V_1$-only Hamiltonian with 10 electrons in 29 orbitals
in Fig. ~\ref{oneqh}(b), and use colormap for the zero-energy edge states
to indicate the corresponding weight in the one-quasihole state.

\begin{figure}
\includegraphics[width=7cm]{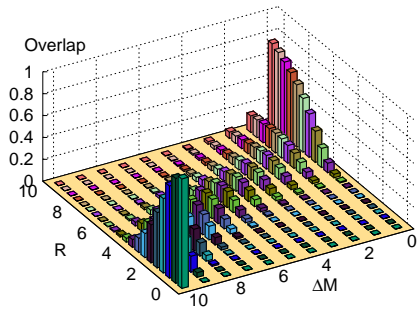}
	\caption{\label{overlapN10} The overlap between the one-quasihole model wave function and the zero-energy edge states as the distance of the quasihole from the center $R = \vert w \vert$ varies. The relevant states evolves from large $\Delta M = N_e$ to zero while the quasihole moving from the center to the edge of the droplet.}
\end{figure}

To further understand the decomposition of the quasihole state into edge states, we plot the
overlap, as defined in Fig. ~\ref{oneqh}(b), as we change the quasihole location $w = R$
in a 3D bar chart in Fig.~\ref{overlapN10}.
Here, the momentum of the edge states is defined as $\Delta M = M_{tot} - M_0$.
One can see that the edge states with significant weight evolve from
having large $\Delta M = N_e$ to zero, when $R$ increases from the center to the edge of disk.
When $R$ is close to half of the radius, the quasihole state incorporates most edge states.

\begin{figure}
\includegraphics[width=9cm]{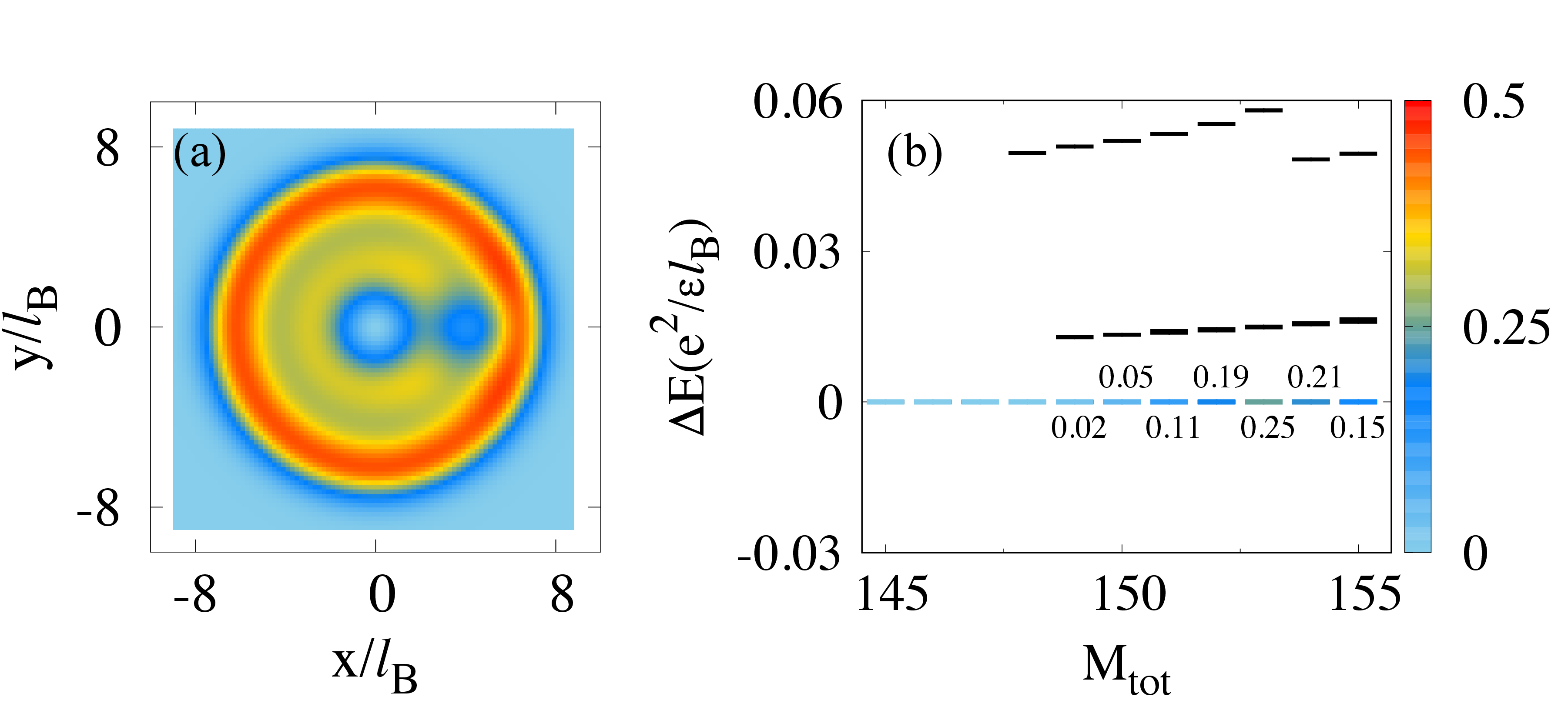}
	\caption{\label{twoqha}(a) The two-quasihole density profile for $10$ electrons and two quasiholes at $w_1 = 0$ and $w_2 = 4l_B$ in case A. (b) The low-lying energy spectrum for $10$ electrons in 30 orbitals with a tip potential at the center. The color illustrates the overlap of these edge states with the corresponding two-quasihole model wave function.}
\end{figure}

For a two-quasihole state, we consider the following two configurations:
(A) with one quasihole at the center $w_1 = 0$ and another at $w_2 = w \neq 0$, and
(B) with two quasiholes symmetrically at $w_1 = -w_2 = w \neq 0$.
Case A is can be straightforwardly derived from the one-quasihole state with an additional zero inserted to the innermost orbital, i.e.,
 \begin{eqnarray}\label{twojack}
 \Psi_{\text{L}}^{\text{2QHA}} &=& |001001001\cdots 1001\rangle - w |010001001\cdots 1001\rangle \nonumber \\
 &+& w^2 |010010001\cdots 1001\rangle + \cdots \nonumber \\
 &+& (-1)^{N_e} w^{N_e}|01001\cdots 10010\rangle.
\end{eqnarray}
Therefore, $\Psi_{\text{L}}^{\text{2QHA}}$ can also be decomposed into a linear combination
of the zero-energy eigenstates of the $V_1$-only Hamiltonian with $3N_e$ orbitals
and one quasihole fixed at the center of the disk.
The density profile of the two-quasihole state with $N_e = 10$, $w_1 = 0$, and $w_2 = 4l_B$ and
its overlap with the zero-energy states are shown in Fig.~\ref{twoqha}.
The zero-energy states with non-zero weight in Fig.~\ref{twoqha}(b) can be regarded as
the edge states of the one-quasihole state with the quasihole fixed at the center.
The distance between two quasiholes is limited by the radius of the disk.
As shown in Fig.~\ref{twoqha}(a), the second quasihole at $w_2 = 4l_B$ can be seen
separated from the quasihole at the center and the edge of the droplet.

\begin{figure}
\includegraphics[width=9cm]{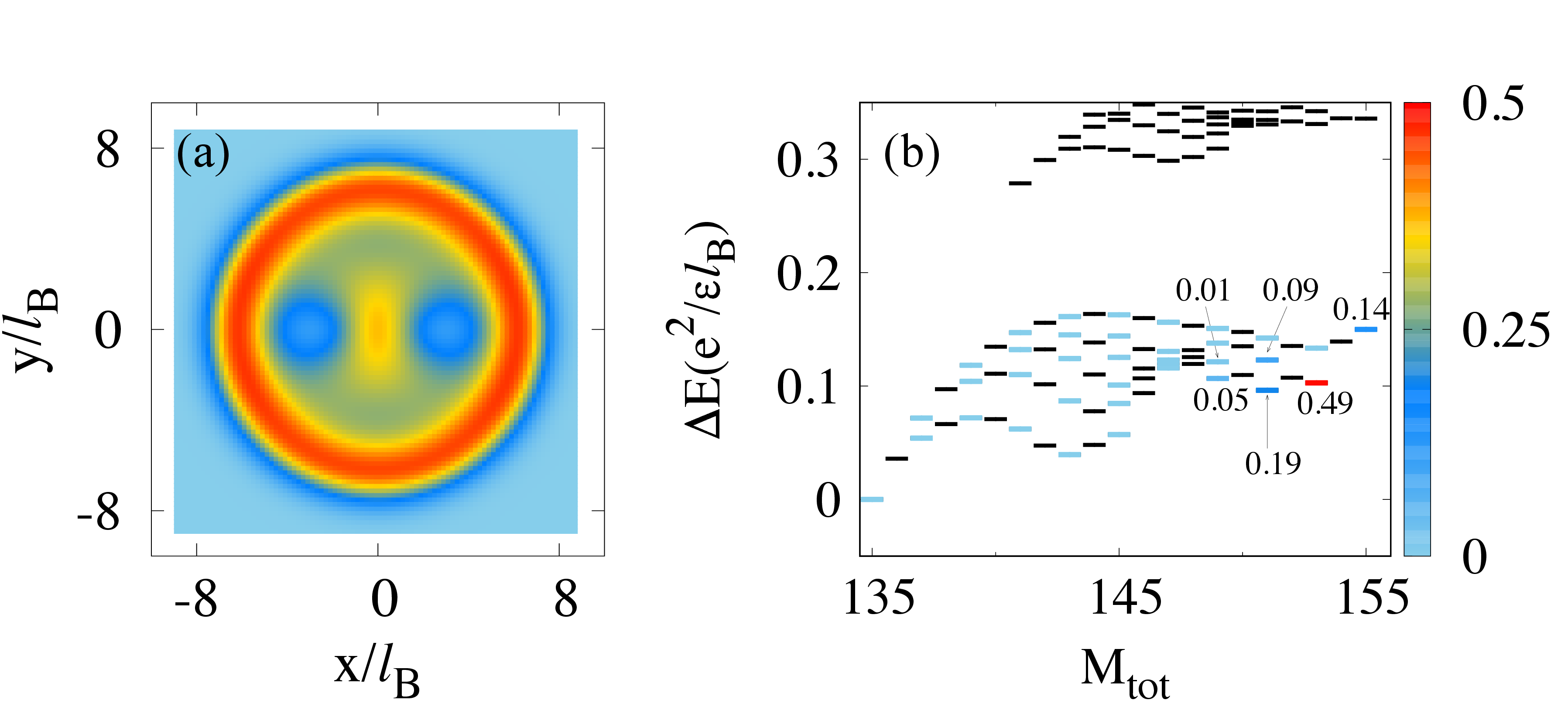}
	\caption{\label{twoqhb}(a) The two-quasihole density profile for $10$ electrons and two quasiholes at $w_1 = -3l_B$ and $w_2 = 3l_B$ in case B. (b) The low-lying energy spectrum for a mixed Hamiltonian (artificially set $V_1 = 10$ in the Coulomb pseudopotentials) in 30 orbitals. The color illustrates the overlap of these edge states with the corresponding two-quasihole model wave function. For the pure $V_1$ interaction, they are all zero-energy edge states.
In lifting the degeneracy of the zero-energy manifold,
we also include in the Hamiltonian a potential from neutralizing background charge at $d = l_B$.~\cite{Hu08}
}
\end{figure}

For the same system size, the distance of the quasiholes can be further separated in case B
as shown in Fig.~\ref{twoqhb}(a).
However, the construction of the two-quasihole state via Jacks is more complicated.
The model wave function with the two quasiholes at $\pm w$ is
\begin{eqnarray}
\Psi_{\text{L}}^{\text{2QHB}} &=& \prod_i(z_i - w) \prod_i(z_i + w) \Psi_{\text{L}} \nonumber \\
&=&  \prod_i(z_i^2 - w^2) \Psi_{\text{L}},
\label{qh2symm}
\end{eqnarray}
the polynomial factor of which can be expanded to be
\begin{eqnarray} \label{qh2expansion}
\prod_i(z_i^2 - w^2)  &=&  \prod_i z_i^2-  w^2 \sum_{j} \prod_{i\neq j}z_i^2
+  w^4 \sum_{j,k} \prod_{i \neq j, k} z_i^2  \nonumber   \\
&&+\cdots + (-1)^{N_e}w^{2N_e}.
\end{eqnarray}
Following the same procedure in the one-quasihole case, we find that
the involved edge states are all zero-energy states for the system with $3N_e$ orbitals and
a total angular momentum $M_{tot} \in [M_0,M_0+2N_e]$.
The symmetrical arrangement of the quasiholes further rules out the contribution
from the states with odd $\Delta M$.
As in the one-quasihole case, the increase of $\vert w \vert$ leads to increasing weight in the small-momentum subspace.
However, each zero-energy momentum subspace can now be degenerate.
The degeneracy prevents us to construct the wave function $\Psi_{\text{L}}^{\text{2QHB}}$
by a single Jack in each subspace.
For $10$ electrons, e.g., the degeneracies are $1,1,2,2,3,3,4,4,5,5,6,5,5,4,4,3,3,2,2,1,1$
for $M_{tot} \in [135,155]$, as shown in Fig.~\ref{twoqhb}.  Those degeneracies could be obtained by counting the number of root configurations of the Jacks for $10$ electrons in $30$ orbitals. In particular, the degeneracy is the largest at $\Delta M = N_e$ and is the same
for $\Delta M = m$ and $2N_e - m$.
In order to expand $\Psi_{\text{L}}^{\text{2QHB}}$ by edge states in a large system,
we need to know how to map the polynomials in Eq.~(\ref{qh2expansion})
to linear combinations of edge Jacks, which we elaborate in Appendix~\ref{sec:app_A}.

\section{fractional statistics}
\label{sec:fs}
\begin{figure}
    \includegraphics[width=7cm]{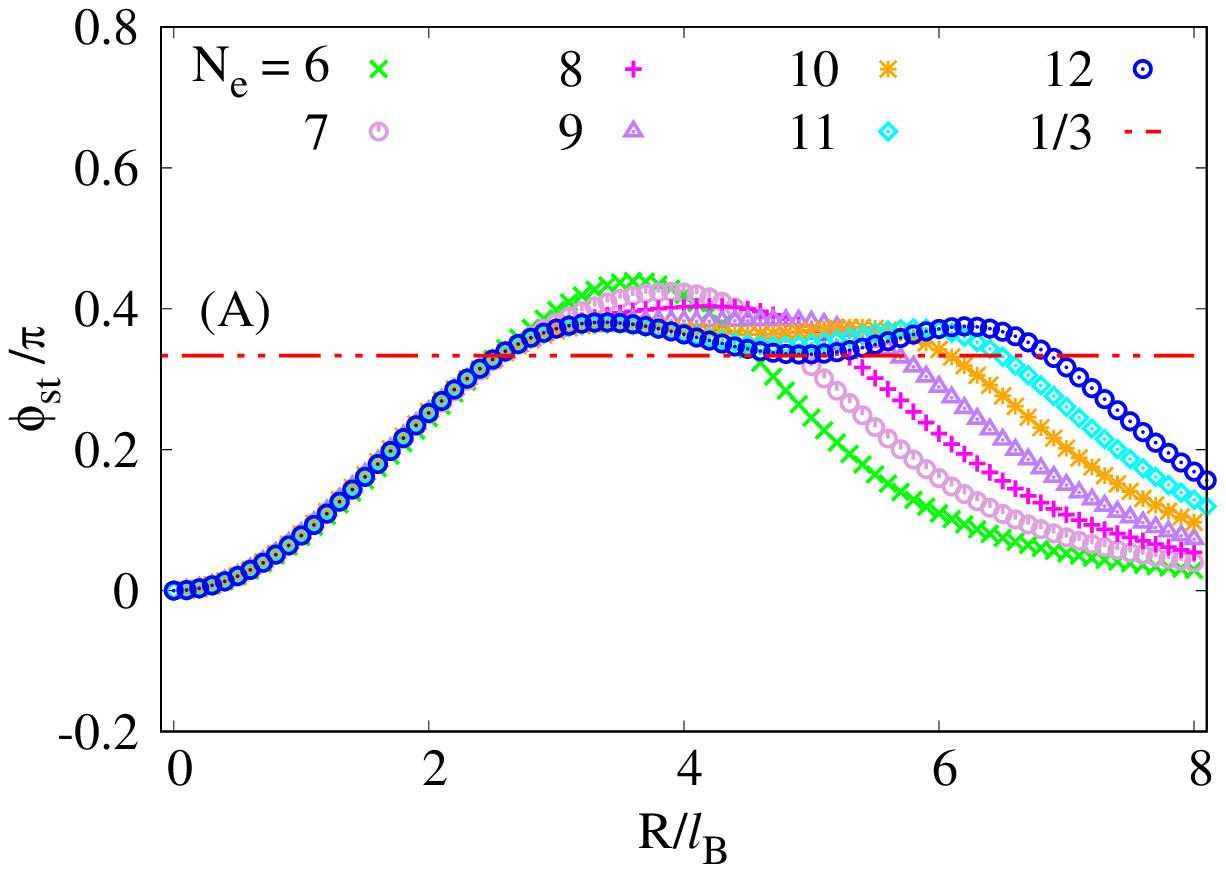}
	\includegraphics[width=7cm]{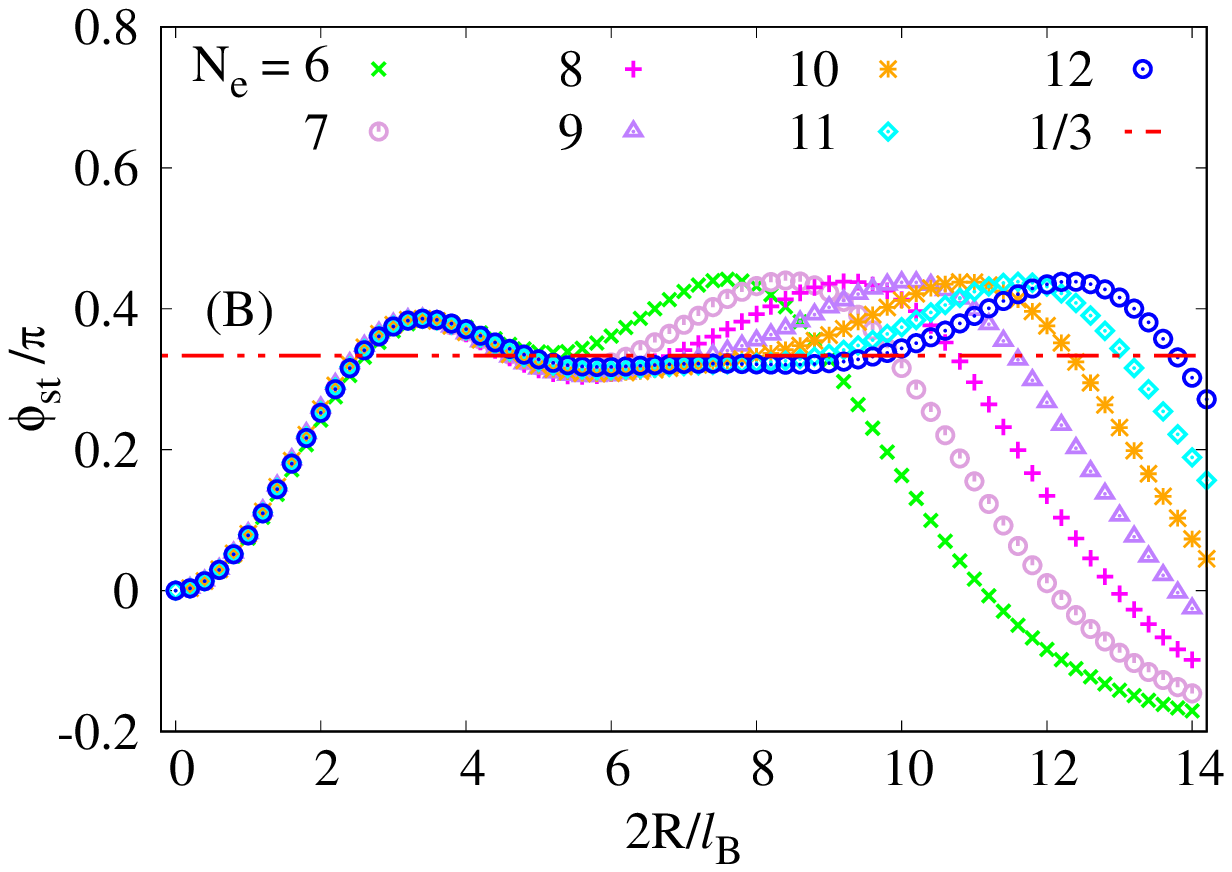}
	\caption{\label{Berryphase}The statistical phase of the Laughlin quasiholes as a function of their distance in both case A (upper panel, $w_1 = 0$ and $w_2 = R$) and case B (lower panel, $w_1 = -R$ and $w_2 = R$). The results for the larger systems with $11-12$ electrons are obtained from the combination of Jack polynomials. Well developed value of $\phi_{st} = \pi/3$ appears when the distance between quasiholes is larger than $5.0l_B$ and when the quasiholes are not too close to the edge.}
\end{figure}
With the exact one-quasihole and two-quasihole wave functions, we can verify the fractional statistics via the Berry phase calculation when the system is adiabatically rotated by a full circle. 
In Eqs.~(\ref{onejack})-(\ref{qh2symm}), the rotation is achieved by letting $\omega \rightarrow \omega e^{i \theta}$ and varying $\theta$ in the wave functions $\Psi(\theta)$, such that the Berry phase is defined as 
\begin{equation}
\phi_B(\theta) = i \oint_{\theta} \langle \Psi(\theta)|\partial_{\theta}|\Psi(\theta)\rangle.
\end{equation} 
The rotation could alternatively be generated by the angular momentum component $\hat{L}_z$ as 
\begin{eqnarray}
|\Psi(\theta + \delta \theta)\rangle &=& \exp(-i\hat{L}_z\delta\theta/\hbar) |\Psi(\theta)\rangle \nonumber \\
&\simeq& (1-i\hat{L}_z\delta\theta/\hbar) |\Psi(\theta)\rangle.
\end{eqnarray} 
Therefore, we have 
\begin{equation}
\phi_B(\theta) = \frac{1}{\hbar} \oint_{\theta} \langle \Psi(\theta)|\hat{L}_z|\Psi(\theta)\rangle  = \frac{2\pi}{\hbar} \langle \hat{L}_z \rangle.
\end{equation}
In our setup, therefore, the mean angular momentum $\langle \hat{L}_z \rangle$ can be related to the fractional statistics of the quasihole. Equivalently, one can calculate the mean square radius $\langle r^2 \rangle$ for the density distribution, because the radius of the $m$th Landau orbital (with angular momentum $m\hbar$) is simply $r = \sqrt{2m}l_B$ in the disk geometry.~\cite{Carusotto18,Carusotto19,Lewenstein20} 

As a result, the mutual braiding statistical phase of two quasiholes can be expressed by the Berry phase of the two-quasihole state subtracting the Berry phases for the two single-quasihole states, 
\begin{equation}
\label{eq:stat}
 \phi_{st} = \frac{\pi}{\hbar} \left (\langle \hat{L}_z \rangle^{\text{2QH}} - \langle \hat{L}_z \rangle_1^{\text{1QH}} - \langle \hat{L}_z \rangle_2^{\text{1QH}} + M_0 \right ) 
\end{equation}
where $\langle \hat{L}_z \rangle_1^{\text{1QH}}$ and $\langle \hat{L}_z \rangle_2^{\text{1QH}}$ are the mean angular momentum for the one-quasihole states with a quasihole at $w$ and 0 for case A, or $\pm w$ for case B. 
The total angular momentum $M_0$ of the Laughlin state is included in Eq.~(\ref{eq:stat}) to compensate for the contributions from the constituent electrons. 
In Fig.~\ref{Berryphase}, we plot the statistical phase as a function of the distance of quasiholes for the two cases. The results are obtained from numerical diagonalizing the model Hamiltonian with $\delta$-function impurity potential for $6-10$ electrons and from the Jack polynomial contruction for $N_e > 10$ electrons. 
The expected value of $\phi_{st} = \pi/3$ for a Laughlin quasihole can be anticipated when (1) the two quasiholes are not too close to each other, and (2) neither quasihole is close to the edge of the droplet. 
The plateau of $\phi_{st} = \pi/3$ is well developed in case B, in which the distance between quasiholes doubles that in case A, while the influence of the edge is the same.  
The plateau starts from a quasihole distance of about $5.0l_B$, which is approximately the diameter of a Laughlin quasihole.~\cite{WuYL} 
This implies that quasiholes can be thought of as independent of each other, as long as they are not overlapping with each other. 
Fig.~\ref{Berryphase} further suggests that when the two quasiholes overlap with each other, their interaction, as reflected in the correction to $\phi_{st}$, is independent of the system size, hence local.

\section{The quench evolution of the two-quasihole state}
\label{sec:quench}

In previous sections we have demonstrated that one- and two-quasihole states of a Laughlin droplet can be constructed by Jack polynomials, which allows us to set up larger systems with two anyons that are sufficiently far from each other and from the edge, such that their mutual fractional statistics can be computed. 
In this section we explore the dynamical evolution of the quasihole states after a quantum quench. 
The exploration of such a quench protocol is motivated by the implementation of topological quantum computing, which needs the braiding of anyons. 
In our setup the system starts with quasiholes localized in a properly designed trapping potential, which is removed or altered at a certain time, say, $t = 0$. 
In the ideal case one quasihole can move around another while still maintaining their local density profile. 
This scheme may be advantageous in realizing braiding, e.g., in GaAs-based systems, in which the two-dimensional electron gas is far from the surface gates.
It might also be the preferred protocol once fractional quantum Hall systems are realized in cold atomic~\cite{Cooper, Gemelke, Miyake, Aide, He, Tobias} or photonic~\cite{LLphoton, photonFQH, Zhou} systems in synthetic gauge field.

\begin{figure}
 	\includegraphics[width=8cm]{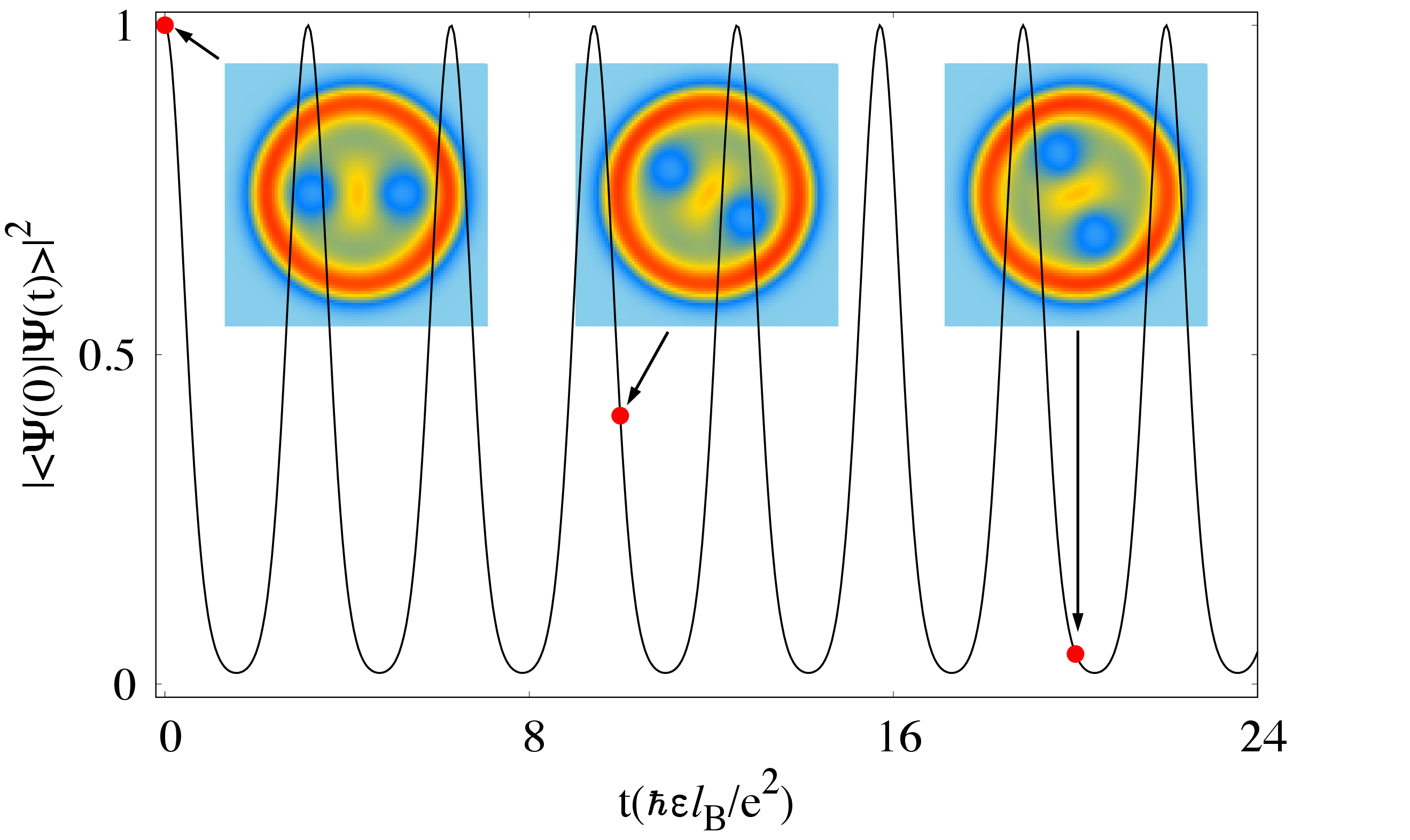}
 	\caption{\label{qhbulk}Time evolution of the wave function overlap $\vert \langle \Psi_{\text{L}}^{\text{2QHB}}(t = 0) \vert \Psi_{\text{L}}^{\text{2QHB}}(t) \rangle \vert^2$ for the two-quasihole state in case B with $10$ electrons in $30$ orbitals. The rotation is driven by an external potential $V(r) = \alpha r^2/2$ with $\alpha = 1.0$.}
\end{figure}

In the first realization we show that the rotation of a quasihole around the other can be realized by a quench protocol, in which a harmonic potential can introduce an angular velocity. 
The harmonic potential, in the form of $V(r) = \alpha r^2/2$ in real space, realizes a linear potential in the momentum space, which introduces a linear dispersion for the edge states. 
We start, in the protocol, with a $V_1$ Hamiltonian for the $\nu = 1/3$ Laughlin liquid and two tip potentials that trap two quasiholes  at $\pm w$. 
In the edge Jack polynomial expansion of the quasihole state as in Eq.(\ref{qh2expansion}), each term has a fixed angular momentum, which is inversely linear in the power of $\omega$. 
At $t = 0$, we turn off the tip potentials and, instead, turn on the harmonic potential with $\alpha = 1.0$.  
After the quench, the system evolves as $|\Psi(t)\rangle = e^{-i\hat{H}t/\hbar}|\Psi(0)\rangle$, 
in which the reduced Hamiltonian contains the potential only. 
Therefore, the dynamical phase difference of these terms can be absorbed by the corresponding polynomials of $\omega$, leading to a pure rotation. 
Fig.~\ref{qhbulk} shows the evolution of the density profile of $\Psi_{\text{L}}^{\text{2QHB}}(t)$ and the overlap of the wave function with $\Psi_{\text{L}}^{\text{2QHB}}(t = 0)$  for $10$ electrons in $30$ orbitals.
We find that the overlap shows regular oscillations, and the density profile is stable in time. 
The period of the oscillations is $\pi$, which corresponds to the exchange of two quasiholes, for which the wave function picks up a negative sign only.

\begin{figure}
 	\includegraphics[width=9cm]{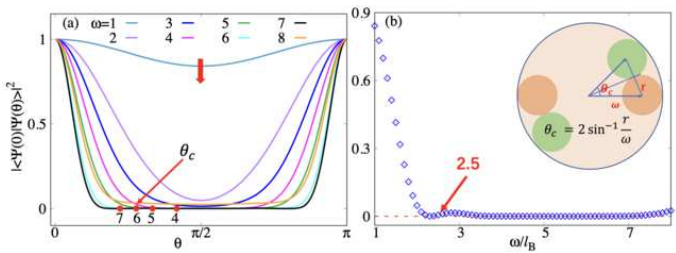}
 	\caption{\label{2qhzero}(a) The two-quasihole wave function overlap $\vert \langle \Psi_{\text{L}}^{\text{2QHB}}(\theta = 0) \vert \Psi_{\text{L}}^{\text{2QHB}}(\theta) \rangle \vert^2$ in a period for $12$ electrons in $36$ orbitals for different $\omega$'s. (b) The overlap at $\theta = \pi/2$ as the quasihole location $\omega$ varies. The inset illustrates the characteristic angle at which the overlap of the quasiholes  with those at $t = 0$ vanishes.}
\end{figure}

The minima in the time-dependent overlap is wider than the maxima in Fig.~\ref{qhbulk}. 
This has an interesting origin and can be used to estimate the size of quasiholes. 
As we have discussed, the time evolution of the wave function is known analytically:
\begin{equation}
\Psi_{\text{L}}^{\text{2QHB}}(\theta)  = \prod_i(z_i^2 - w^2 e^{2 i \theta}) \Psi_{\text{L}},
\end{equation}
where $\theta = \alpha t$.
In Fig. ~\ref{2qhzero}(a) we plot the overlap in a period $\theta \in [0, \pi]$ with different $\omega$ for a system with $12$ electrons in $36$ orbitals. 
For a small $\omega$, the oscillation is sinusoidal, and a nonzero minimum appears at $\theta = \pi/2$ (indicated by the vertical arrow), when the quasiholes are rotated to $\pm i \omega$.
As $\omega$ increases, the overlap at $\theta = \pi/2$ decreases and vanishes at $\omega \geq 3 l_B$, whose value is close to the radius of a quasihole $r \approx 2.5 l_B$, which indicates that the nonzero value of the minimum may come from the spatial overlap of the quasiholes. 
Further, the overlap deviates from zero, again, at $w = 8 l_B$, presumably when the quasiholes touch the rim of the droplet. 
The continuous dependence of the overlap at $\theta = \pi/2$ on $w$ is shown in Fig.~\ref{2qhzero}(b). 
To understand the curves in Fig.~\ref{2qhzero} and their connection to the quasihole size, we model quasiholes by rigid disk of radius $r$. 
As illustrated in the inset of Fig.~\ref{2qhzero}(b), if the wave function overlap vanishes when quasiholes no longer overlap with each other (for sufficiently large $w$), there exists a characteristic angle of rotation
\begin{equation}
\theta_c = 2 \sin^{-1}\frac{r}{\omega}.
\end{equation}
In Fig. ~\ref{2qhzero}(a) we mark the angles for $w = 4,5,6,7$ by red dots on the corresponding curves and find that they explain well the flat bottom of the overlap curves. 

\begin{figure}
 	\includegraphics[width=8cm]{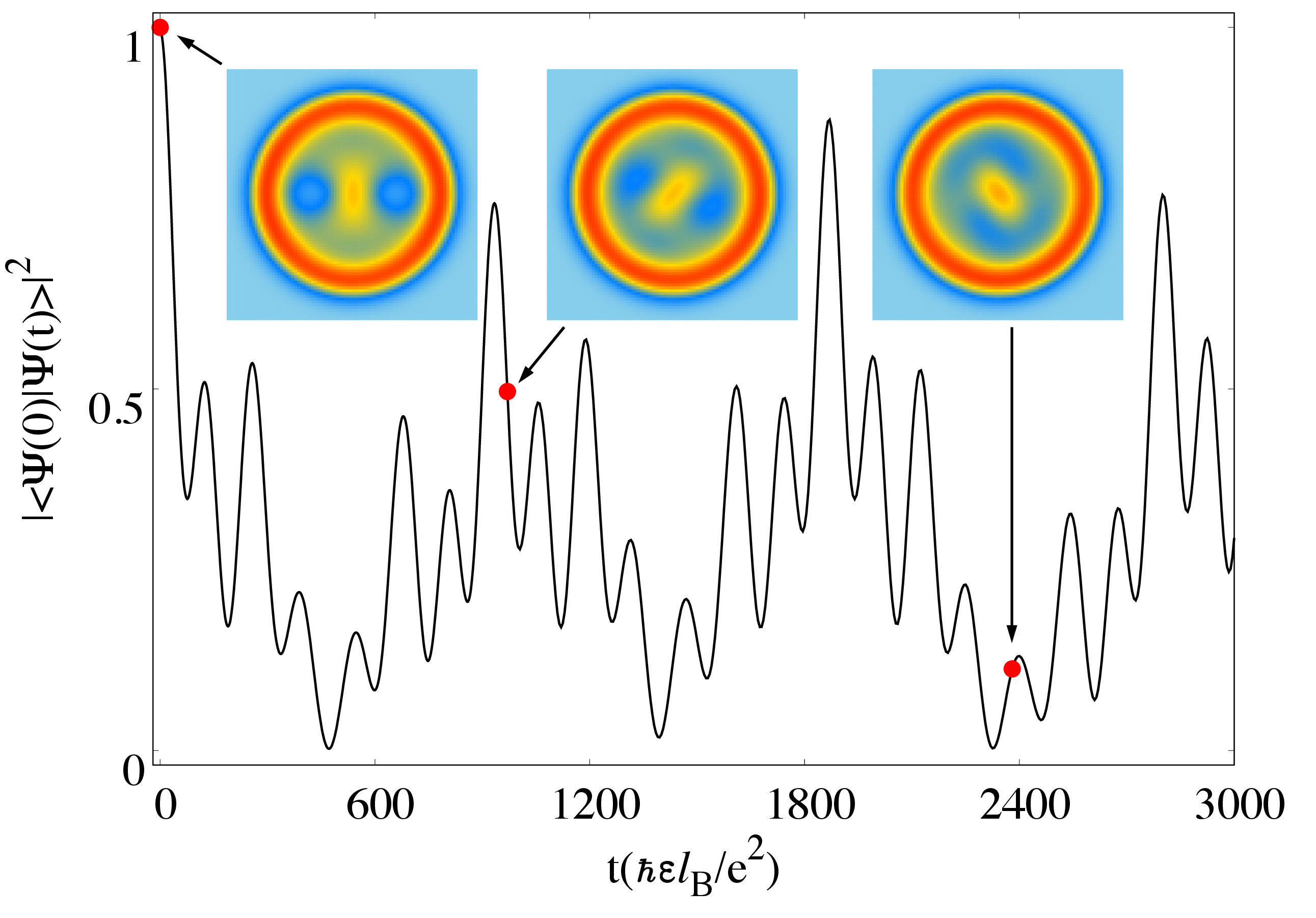}
 	\caption{\label{qhedge} Time evolution of the wave function overlap $\vert \langle \Psi_{\text{L}}^{\text{2QHB}}(t = 0) \vert \Psi_{\text{L}}^{\text{2QHB}}(t) \rangle \vert^2$ for the two-quasihole system in case B with long-range interaction and confinement potential, which drive the quasihole rotation and blurring.}
\end{figure}

A more realistic situation can involve some long-range interaction and a neutralizing confining potential.  We consider the following mixed Hamiltonian: 
\begin{eqnarray}
H = \lambda_1 V_1 + \lambda_2 H_C,
\end{eqnarray}
where $V_1$ is the model Hamiltonian and $H_C$ is the Coulomb Hamiltonian 
\begin{equation}
 H_C = \sum_{m_1 m_2 m_3 m_4}V_{1234}C_{m_1}^{\dagger} C_{m_2}^{\dagger} C_{m_3}C_{m_4} + \sum_m U_m C_m^{\dagger} C_m.
\end{equation}
which contains the Coulomb interaction among electrons and between electrons and background charges~\cite{Wan03}.
Here $\lambda_{1,2}$ are parameters that we tune to adjust the bandwidth of the low-energy edge-state manifold and its gap to the higher-energy states. 
As in Fig.~\ref{twoqhb}, we choose $\lambda_1 = 10$ and $\lambda_2 = 1$. 
In general, the direct calculation of the matrix exponential in $|\Psi(t)\rangle = e^{-i\hat{H}t/\hbar}|\Psi(0)\rangle$ is time and memory consuming for large systems.  For example, the two-quasihole state for $12$-electron has a full dimension of 1,251,677,700 in $36$ orbitals. 
The situation is significantly simpler when we assume that the quasiholes are prepared in the model wave function in Eq.~(\ref{qh2symm}), which can be written as the linear combination of a few edge states, as demonstrated in Fig.~\ref{twoqhb}. 
These edge states are protected by an energy gap, which origins from a hard wall edge potential due to the choice of the fixed number of orbitals. 
If we assume that the quasihole state do not leak outside the low-energy manifold, the time evolution can be calculated only in the truncated edge space. 
For the $12$-electron two-quasihole state, the dimension of the reduced Hamiltonian is merely $49$ and we have 
\begin{eqnarray}
\label{eq:expansion}
|\Psi(t)\rangle = \sum_i c_i e^{-i E_i t/\hbar}|\phi_i\rangle,
 \end{eqnarray}
where $E_i$ and $|\phi_i\rangle$ are the energy and wave function of the relevant edge states, and we have $c_i = \langle \phi_i|\Psi(0)\rangle$.  
However, due to the presence of Coulomb interaction, the dispersion of the edge modes is no longer linear, as is evident in Fig.~\ref{twoqhb}. 
In Fig.~\ref{qhedge} we plot the wave function overlap $\vert \langle \Psi_{\text{L}}^{\text{2QHB}}(t = 0) \vert \Psi(t) \rangle \vert^2$ as a function of time. 
The curve oscillates with multiple periods. 
In particular, we can identify a period of $T \approx 940$, at the integral multiple of which 
the overlap is about 0.75 or more. 
A smaller period, which is about $T/7$ is also visible in the overlap. 
Meanwhile, the plots of density profiles indicate that the nonlinear dispersion of the edge mode deforms the shape of the quasihole in a long time evolution. 
The smear-out of the quasiholes is significant when the overlap is small. 

 \begin{figure}
 	\includegraphics[width=8cm]{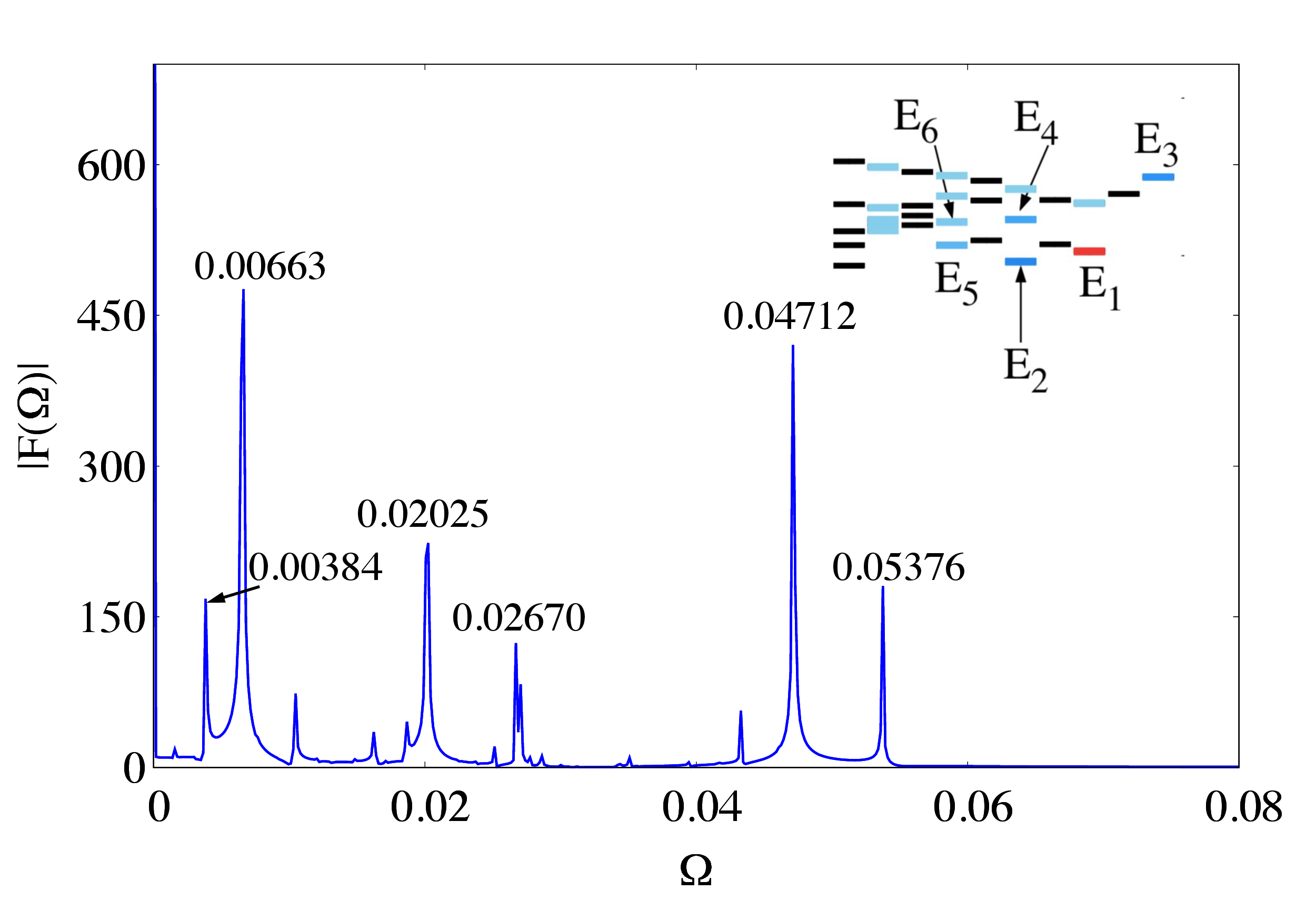}
 	\caption{\label{fft} The Fourier frequency spectrum of the wave function overlap in Fig.~\ref{qhedge}. The inset shows the relevant part of the energy spectrum in Fig.~\ref{twoqhb}(b). The edge states labelled by $E_1, E_2, \dots, E_6$ are those that have appreciable overlap with the two-quasihole wave function.}
 \end{figure}

The oscillations in the overlap can be understood as follows. 
According to Fig.~\ref{twoqhb}, the expansion of the time-dependent wave function in Eq.~(\ref{eq:expansion}) is dominated by a few edge states (only six states have a weight larger than 0.01). 
We can calculate the overlap to be 
\begin{align}
& \vert \langle \Psi(0) \vert \Psi(t) \rangle \vert^2 = \left \vert \sum_{i} \vert c_i \vert^2 e^{-i E_i t/\hbar} \right \vert^2 \nonumber \\ 
=& \sum_i \vert c_i \vert^4 + \sum_{i < j} 2 \vert c_i \vert^2 \vert c_j \vert^2 \cos \frac{(E_i - E_j) t}{\hbar}.
\end{align}
Therefore, the time-dependent overlap is dominated by the energy differences of two edge states, both with significant probability amplitudes. To show this, we calculate the Fourier transformation of the data in Fig.~\ref{qhedge} and plot the spectral weight in Fig.~\ref{fft}. 
In the inset we copy the part of the low-energy spectrum in Fig.~\ref{twoqhb}(b) that includes the edge states with appreciable weight. 
In particular, we label the six most important components by $E_1, E_2, \dots, E_6$, according to the order of their weights. 
The peaks in the Fourier spectrum are in good agreement with the energy differences among them, as listed in Table~\ref{deltaE}. 
Specifically, we have $\hbar \Omega_{12} = E_1 - E_2 = 0.00655$, which corresponds to a period $2 \pi/\Omega_{12} \approx 959$; this explains the period of the major revivals of the overlap. 
The subleading peak corresponds to $\hbar \Omega_{31} = E_3 - E_1 = 0.04717$, which is 7.2 times larger than $\hbar \Omega_{12}$; this explains the shorter period that appears in Fig.~\ref{qhedge}. 
The third leading peak is located at $\hbar \Omega_{41} = E_4 - E_1 = 0.02016$; the corresponding period is not easy to detect by visual examination of Fig.~\ref{qhedge} as it is close to $\hbar \Omega_{31}/ 2$ and roughly $3 \hbar \Omega_{12}$. 
The leading seven contributions, resulting in the highest peaks in Fig.~\ref{fft} (notice $E_3 - E_4 \approx E_4 - E_2$), are underlined in Table~\ref{deltaE}.

\section{Discussions and Conclusions}
\label{sec:dis}
In summary, we construct one-quasihole and two-quasihole wave functions for a Laughlin quantum Hall state via numerically diagonalizing the model Hamiltonian with an impurity potential or the combination of the edge Jacks. In the Jack polynomial construction, we identify a universal mapping between the edge Jacks and the  Jastrow term of the Laughlin wave function multiplied by symmetric polynomials, which allows us to approach systems with up to $12$ electrons.
These quasihole states lives entirely in the degenerate zero-energy Hilbert space with a hard-wall confinement; this fact tremendously reduces the dimension of the Hilbert space in the time evolution. 

The fractional statistics of the quasiholes can be computed through the Berry phase or, effectively, the mean angular momentum of those states, as long as the quasiholes are not overlapping with each other and with the edge of the FQH droplet. 
The novel technique with Jack polynomials provides us a reasonably wide range of quasihole location, in which we can observe a plateau of the expected anyon statistics. 

Based on a quench protocol, we demonstrate that quasiholes in the bulk can be driven by a harmonic potential or, in the more realistic case, by Coulomb interaction and background charge confinement to braid around each other. 
This reveals interesting possibilities of manipulating anyons by dynamical quench in numerical simulations and, possibly, in future experiments toward the realization of topological quantum computation with anyons in FQH liquids.

\begin{table}
 \begin{center}
  \caption{\label{deltaE}The table of absolute value of the energy difference $|E_i-E_j|$ for the edge states marked in the inset of Fig.~\ref{fft}. The underlined energy differences are consistent to the frequency peaks in the Fourier spectrum of the overlap $\vert \langle \Psi(0) \vert \Psi(t) \rangle \vert^2$ in Fig.~\ref{qhedge}.}
  \begin{tabular}{c c c c c c c} 
  \hline \hline
    & $E_1$ & $E_2$ & $E_3$ &$E_4$ &$E_5$ &$E_6$ \\
   \hline
   $E_1$\hspace{0.2cm}  & 0&  &  &  &\\
   $E_2$\hspace{0.2cm}  &  \underline{0.00655}& 0&  &  &\\
   $E_3$\hspace{0.2cm}  &   \underline{0.04717} &  \underline{0.05373}& 0&  &\\
   $E_4$\hspace{0.2cm}  &   \underline{0.02016}&  \underline{0.02672} &  \underline{0.02701} &  0&\\
   $E_5$\hspace{0.2cm}  &   \underline{0.00387}& 0.01042 & 0.04330 & 0.01629 & 0&\\
   $E_6$\hspace{0.2cm}  &  0.01860& 0.02516 & 0.02857  & 0.00156 & 0.01473 &  0\\
  \hline \hline
  \end{tabular}
 \end{center}
\end{table}

\acknowledgments
This work at Chongqing University was supported by National Natural Science Foundation of China No.11974064,  the Chongqing Research Program of Basic Research and Frontier Technology No.cstc2021jcyj-msxmX0081, Chongqing Talents: Exceptional Young Talents Project No.cstc2021ycjh-bgzxm0147, and the Fundamental Research Funds for the Central Universities No. 2020CDJQY-Z003. N.J. was supported by National Natural Science Foundation of China No.12104075. X.W. was supported by the Strategic Priority Research Program of Chinese Academy of Sciences through Grant No. XDB28000000.

\appendix

\section{Mapping Two-Quasihole States to Jack Polynomials}
\label{sec:app_A}

In Sec.~\ref{sec:qh} we discussed the principles and difficulties of constructing a
two-quasihole state with a symmetric configuration as in case B.
The apparent difficulty is that each term in Eq.~(\ref{qh2expansion}),
the polynomial expansion of the model wave function,
involves multiple zero-energy Jacks in the same momentum subspace.
We need to find a series of matrices in these subspaces that can map the
polynomials to the corresponding Jacks.

For simplicity, let us take a  $4$-electron Laughlin state as a concrete example.
Multiplied by the Jastrow term $\prod_{i<j}(z_i - z_j)^3$, the first term on the right-hand side of Eq.~(\ref{qh2expansion}) corresponds uniquely to the Jack $|\red{00}1001001001\rangle$,
which lies in the momentum subspace with $\Delta M = 2N_e=8$.
The second term $\sum_{j} \prod_{i\neq j}z_i^2$, however, corresponds to the Jack
$|10\red{00}01001001\rangle$ and its descendant from the squeeze rule
$|\red{0}1001\red{0}001001\rangle$.
They both have $\Delta M =6$ and span a two-dimensional degenerate subspace.
In the same manner, the third term in Eq.~(\ref{qh2expansion}) can be expanded by $|10010\red{00}01001\rangle$ and its two descendants $|100\red{0}1001\red{0}001\rangle$ and $|\red{0}1001001001\red{0}\rangle$, which result in a three-fold degeneracy in the $\Delta M = 4$ subspace.
Now the question is how to combine degenerate Jacks to construct all the terms
in Eq.~(\ref{qh2expansion}) multiplied by the Jastrow factor.
In other words, we need to calculate the coefficient $c_{\mu}$ in the expression
 \begin{equation} \label{expand}
\pmb{m}_{\Delta M}\prod_{i<j}(z_i - z_j)^3 = \sum _{\mu \preceq \Omega(\Delta M)}c_{\mu} J_{\mu}.
\end{equation}
where $\pmb{m}_{\Delta M}$ is the polynomial in Eq.~(\ref{qh2expansion}) with degree $\Delta M$.
The summation of $J_{\mu}$ contains all the Jacks whose root configurations descend from
$\Omega(\Delta M)$, which we denote for the Jack $|1001\cdots10010\red{00}01001\cdots\rangle$
with $\Delta M/2$ electrons (or 1s) to the right of the two inserted quasiholes (or 0s, labelled in red).
The coefficients $\{c_{\mu}\}$ need to be determined with the nonorthogonality
among $J_{\Omega(\Delta M)}$ and its descendants in mind.
The details resemble the monomial expansion of an arbitrary monomial multiplied by the Laughlin state,
which we need in the construction of edge excitations; in the latter case, however, the summation is taken without restricting the orbital number.~\cite{LeeJack}
Therefore, we have here a simpler situation in which the two-quasihole state lives
in fixed $3N_e$ orbitals, so the degeneracy is smaller.
Taking the 4-electron system as an example, we determine $\{c_{\mu}\}$ for $\Delta M=2$
by solving the following identity
$$\begin{pmatrix}
\text{Sym}[z_iz_j] \\
\text{Sym}[z_i^2]
\end{pmatrix} J_{\Omega(0)} =
A_2\begin{pmatrix}
|100100010010\rangle\\
|100100100001\rangle
\end{pmatrix}$$
where $J_{\Omega(0)} \equiv \vert 1001001001 \rangle= \prod_{i<j}(z_i - z_j)^3$ is the Laughlin state and $A_2$ is a $2 \times 2$ lower triangular matrix in the following form
\begin{equation}
A_2=\begin{pmatrix}
1 & 0 \\
-\frac{6}{5}&1   \end{pmatrix}.
\end{equation}
For $\Delta M=4$, we need to solve
$$\begin{pmatrix}
\text{Sym}[z_iz_jz_kz_l] \\
\text{Sym}[z_i^2z_jz_k] \\
\text{Sym}[z_i^2z_j^2]
\end{pmatrix} J_{\Omega(0)}  =
A_3\begin{pmatrix}
|010010010010\rangle\\
|100010010001\rangle\\
|100100001001\rangle
\end{pmatrix},$$
which yields a $3 \times 3$ lower triangle matrix
\begin{equation}\renewcommand*{\arraystretch}{1.3}
A_3=\begin{pmatrix}
1 & 0 &0\\
\frac{-36}{11}&1&0\\
\frac{27}{22}  &\frac{-6}{5}&1 \end{pmatrix}.
\end{equation}
For $\Delta M = 6$, we need, again, a $2 \times 2$ matrix, which turns out to be identical to $A_2$
and satisfies
$$\begin{pmatrix}
\text{Sym}[z_i^2z_j^2z_kz_l] \\
\text{Sym}[z_i^2z_j^2z_k^2]
\end{pmatrix} J_{\Omega(0)}  =
A_2\begin{pmatrix}
|010010001001\rangle\\
|100001001001\rangle
\end{pmatrix}.$$
These are the matrices needed for the 4-electron system with 12 orbitals, whose
degeneracies in the zero-energy subspace for $M_{tot} \in [18, 26]$
are $1,1,2,2,3,2,2,1,1$, respectively.
But in a system with more than 5 electrons, the degeneracy for $\Delta M = 6$
becomes 4, and the corresponding matrix is
\begin{equation}\renewcommand*{\arraystretch}{1.3}
 A_4=\begin{pmatrix}
1 & 0 &0&0\\
\frac{-90}{17}&1&0&0\\
\frac{810}{119}  &\frac{-36}{11}&1&0\\
\frac{-1620}{1309} &\frac{27}{22}&\frac{-6}{5}&1\end{pmatrix}.
 \end{equation}

Interestingly, the $n$-dimensional matrix $A_n$ is embedded as the lower right submatrix
of the $(n+1)$-dimensional matrix $A_{n+1}$.
We find these matrices are independent of the system size and,
as shown in the 4-electron case, the same for $\Delta M = m$ and $2N_e - m$.
Since the largest system we calculate in this work contains 12 electrons,
all we need to present is the following $7 \times 7$ matrix
for $\Delta M = 12$ subspace:
\begin{equation}\renewcommand*{\arraystretch}{1.3}
A_7=\begin{pmatrix}
  1&0&0&0&0&0&0\\
        -\frac{396}{35}&1&0&0&0&0&0\\
        \frac{2673}{56}&-\frac{270}{29}&1&0&0&0&0\\
        -\frac{2673}{29}&\frac{11340}{377}&-\frac{168}{23}&1&0&0&0\\
        \frac{120285}{1508}&-\frac{340200}{8671}&\frac{378}{23}&-\frac{90}{17}&1&0&0\\
        -\frac{216513}{8671}&\frac{153090}{8671}&-\frac{4536}{391}&\frac{810}{119}&-\frac{36}{11}&1&0\\
        \frac{216513}{173420}&-\frac{183708}{147407}&\frac{486}{391}&-\frac{1620}{1309}&\frac{27}{22}&-\frac{6}{5}&1\end{pmatrix}.
\end{equation}

When two quasiholes are arranged in an asymmetric configuration, i.e., $w_1\neq w_2$,
the odd $\Delta M$ subspaces are also involved.
The matrices $B_n$ for these momentum space are, again, symmetric around
$M_{tot} = M_0 + N_e$.
It still holds that $B_n$ is embedded as the lower right submatrix in $B_{n + 1}$.
So, we only need to present the largest matrix $B_6$ for $\Delta M = 11$ and $N_e = 12$:
\begin{equation}
\begingroup
\renewcommand*{\arraystretch}{1.5}
B_6=\begin{pmatrix}
 1&0&0&0&0&0\\
        -\frac{165}{16}&1&0&0&0&0\\
        \frac{4455}{116}&-\frac{108}{13}&1&0&0&0\\
        -\frac{93555}{1508}&\frac{6804}{299}&-\frac{63}{10}&1&0&0\\
        \frac{340200}{8671}&-\frac{6804}{299}&\frac{189}{17}&-\frac{30}{7}&1&0\\
        -\frac{505197}{69368}&\frac{30618}{5083}&-\frac{81}{ 17}&\frac{270}{77}&-\frac{9}{4}&1\end{pmatrix}.
\endgroup
\end{equation}

Finally, we emphasize that the lowest row in each matrix corresponds to the coefficients $\{c_{\mu}\}$ in Eq.~(\ref{expand}).
Therefore, we can construct the product of each term in Eq.~(\ref{qh2expansion})
and the Jastrow term by a linear combination of Jack polynomials.
This concludes our construction of the two-quasihole wave function
$\Psi_{\text{L}}^{\text{2QHB}}$ with only $A$ matrix, or in the more general asymmetric case with both $A$ and $B$ matrix.

\end{document}